\documentclass{elsart5}
 \usepackage{graphicx}

\usepackage{amssymb}

\begin{document}

\begin{frontmatter}

\title{Limit to manipulation of qubits due to spontaneous symmetry breaking}
\author[aff1]{Jasper van Wezel}
\author[aff1,aff2]{Jeroen van den Brink\corauthref{cor1}}
\ead{brink@ilorentz.org}
\ead[url]{www.ilorentz.org/$\sim$brink}
\corauth[cor1]{}
\address[aff1]{Institute-Lorentz for Theoretical Physics,  Universiteit  Leiden,\\
P.O. Box 9506, 2300 RA Leiden,The Netherlands}
\address[aff2]{Institute for Molecules and Materials, Radboud Universiteit Nijmegen,\\ 
P.O. Box 9010, 6500 GL Nijmegen, The Netherlands}
%\received{12 June 2005}
%\revised{13 June 2005}
%\accepted{14 June 2005}

\begin{abstract}
We consider a two-spin qubit that is subject to the orderparameter field of a symmetry broken manipulation device. It is shown that the thin spectrum of the manipulation device limits the coherence of the qubit. 
\end{abstract}

%%%%%%%%%use  the \KEY command at the begin of keyword text%%%%%%%%%
\begin{keyword}
\PACS 03.65.Yz\sep 75.10.-b\sep 75.45.+j
\KEY  quantum coherence\sep quantum spin models \sep  symmetry breaking
\end{keyword}
%Please supply one or more relevant PACS-1996 classification codes 
%(http://publish.aps.org/PACS/96pacs.htmland) and about 5 keywords 
%of your own choice for indexing purposes. 
%You can see a list of already used keywords for JMMM at 
%http://authors.elsevier.com/JournalDetail.html?PubID=505704&Precis=KIND

\end{frontmatter}
The  experimental progress in the realization of quantum superpositions --qubits-- is staggering: we have, nowadays, a proliferation of different kinds of spin, charge and  superconducting qubits. In order to use these for quantum computation it is essential that the qubits can be manipulated. Here we consider a two-spin qubit that is manipulated by an external magnetic field. The important point is that we assume that the external field is not just some presupposed classical magnetic field. Instead we take it to be generated by the order parameter --the magnetization-- of a macroscopic quantum magnet. This magnet, which is our manipulation device, is necessarily in a symmetry broken state as it has a finite orderparameter. This implies the existence of a thin spectrum in the manipulation device. We show that precisely this sets a upper bound for the coherence of the qubit that is being manipulated. This limit becomes important when the device is small.

Our analysis is based on the Lieb-Mattis spin Hamiltonian. We have recently shown that in the framework of this Hamiltonian one can derive  a limit to quantum coherence in {\it many-particle} spin qubits~\cite{Wezel05,Wezel06}. This universal limit is due to spontaneous symmetry breaking and the time-scale is $t_{spon} \simeq 2\pi N \hbar / (k_B T)$, given in terms of the number of microscopic degrees of freedom $N$, temperature $T$, and the constants of Planck ($\hbar$) and Boltzmann ($k_B$).
  In the present paper, however, we will consider the many-body spin system as a manipulation device and not as qubit. 

The Hamiltonian of the symmetry broken manipulation device is given by the Lieb-Mattis quantum antiferromagnet~\cite{Lieb62}, with the Hamiltonian
\begin{eqnarray}
H_{LM} = \frac{2J}{N} {\bf S}_A \cdot {\bf S}_B - B (S_A^z - S_B^z).
\end{eqnarray}
The Hamiltonian is defined for a bipartite system with $A$ and $B$ sub-lattices, where ${\bf S}_{A/B}$ is the total spin on the $A/B$ sublattice with $z$-projection $S_{A/B}^z$, and $B$ is the symmetry breaking field, in this case a staggered magnetic field acting on the staggered magnetization $S_A^z - S_B^z$. The particularity of the Lieb-Mattis Hamiltonian is that every spin on sublattice $A$ is interacting with {\it all} spins on sublattice $B$ and vice versa, with interaction strength $2J/N$ (which depends on the number of spins $N$ so that the system is extensive). This very simple Hamiltonian accurately describes symmetry breaking and the related thin spectrum that is encountered in more complicated Hamiltonians, like the nearest neighbor Heisenberg antiferromagnet~\cite{Anderson52,Kaplan90,Kaiser89}.

As an example of a qubit manipulation, we consider the rotation of a two spin qubit from its singlet state into a triplet state. To do this a local staggered magnetic field acting on the qubit is needed. For this we use the orderparameter field of a symmetry broken  antiferromagnet. 

We will now show that from the very moment that the qubit and antiferromagnet are coupled, the manipulation device, because of its intrinsic thin spectrum, starts to decohere the two-spin qubit. The model Hamiltonian describing this process is given by:

\begin{eqnarray}
H=H_{LM} + \Delta {\bf S}_1 \cdot {\bf S}_2 + \frac{\gamma}{N} \left( S_A^z-S_B^z \right) \left( S_1^z-S_2^z \right),
\label{H}
\end{eqnarray}
where we divide $\gamma$ by $N$ to ensure that the spin-spin coupling is of order $J$. In this model $\Delta$ describes energy splitting between the singlet and triplet state of the qubit. If we first take $\Delta$ to be zero, then we can diagonalize the Hamiltonian exactly, and write its eigenfunctions as simple product functions~\cite{Wezel05,Wezel06}:
\begin{eqnarray}
H \left| n, S^z_1, S^z_2 \right> = E\left(n, S^z_1, S^z_2 \right) \left| n, S^z_1, S^z_2 \right>.
\end{eqnarray}
Here $\left| n \right>$ are the eigenfunctions of the symmetry broken Lieb Mattis antiferromagnet and $S^z_1$ and $S^z_2$ are the z-projections of the qubit spins. With  these eigenstates we can now describe the experiment in which the qubit singlet state is turned into the triplet state by the magnetic coupling to the antiferromagnet. First we construct the initial density matrix:
\begin{eqnarray}
\rho_{t<t_0} = \frac{1}{Z} \sum_{n} e^{-\beta E( n)} 
\left| n \right> \otimes    \left| qubit \right> \cdot
\left< qubit \right| \otimes \left< n \right|,
\end{eqnarray}
where $\left|qubit \right> = \frac{1}{\sqrt{2}} \left( \left|\uparrow \downarrow \right> - \left|\downarrow \uparrow \right> \right)$. Then we let this density matrix evolve in time, using the exact time evolution operator:
%
%\begin{eqnarray}
$\rho_{t>t_0} = U \rho_{t<t_0} U^{\dagger}.$
%\end{eqnarray}
%
Finally, we trace away the complete antiferromagnet, since we are interested in the qubit state only, not in the exact state of the manipulation device. Notice that in this case the tracing of the antiferromagnetic states boils down to tracing away the thin spectrum states:
\begin{eqnarray}
\rho^{red}_{t>t_0} = \sum_{\psi} \langle \psi|\rho_{t>t_0}|\psi\rangle 
= \frac{1}{Z} \sum_{n} e^{-\beta E\left(n\right)} e^{-\frac{i}{\hbar} \sqrt{\frac{J}{h}} \frac{\gamma}{4 N} n t}
\end{eqnarray}
This final expression for the coherence of the two-spin qubit state yields the coherence time:
\begin{eqnarray}
t_{coh} \propto \frac{N \hbar}{k_B T} \frac{B}{\gamma}.
\label{eq:t_coh}
\end{eqnarray}
This coherence time-scale thus limits the time available to perform a manipulation on the qubit. Since the decoherence of separate manipulations presumably will have an accumulating adverse effect, this same time-scale also limits the total time that a quantum computer will have to run its calculation, if it uses nanoscopic symmetry broken manipulation machines.

We consider the case with non-zero $\Delta$. When $\Delta$ is large, the singlet will not easily be rotated into a triplet. This limit is not very practical since it also implies that the antiferromagnet will not be able to function as a manipulation device. We therefore consider the limit $\Delta \ll \gamma$.
In this case we can no longer diagonalize Hamiltonian~(\ref{H}) analytically. To study the time dependent decoherence we use the dynamical mean field method described by Allahverdyan e{\it t al.}~\cite{Balian}. First we split the Hamiltonian~(\ref{H}) in as $H = H_{AF} + H_{qubit}$ and then introduce the following meanfield Hamiltonians for the antiferromagnet and qubit 
\begin{eqnarray}
H_{AF}  &=& H_{LM} + \frac{\gamma}{N} \left< S_1^z-S_2^z \right> \left( S_A^z-S_B^z \right)\nonumber \\
H_{qubit} &=& \Delta {\bf S}_1 \cdot {\bf S}_2 + \frac{\gamma}{N} \left< S_A^z-S_B^z \right> \left( S_1^z-S_2^z \right).
\label{Hdec}
\end{eqnarray}
We then use the adiabatic assumption to set $\left< S_A^z-S_B^z \right>$ to its semi-classical value $N/2$ so that we can diagonalize $H_{qubit}$ exactly. The resulting eigenstates can be written in the eigenbasis of the operator $S_1^z-S_2^z$ as:
\begin{eqnarray}
\left|\psi_{qubit}(t)\right> = \sqrt{\frac{1}{2}} \left( \chi_{\uparrow \downarrow}(t) \left| \uparrow \downarrow \right> + \chi_{\downarrow \uparrow }(t) \left| \downarrow \uparrow \right> \right).
\label{vec}
\end{eqnarray}
The time dependence of the components in this eigenstate is given by the time evolution operator which corresponds to $H_{qubit}$. To describe the dynamical behavior of the complete system we now follow Allahverdyan {\it et al.} by writing:
\begin{eqnarray}
\left|\psi (t)\right> &=& \sqrt{\frac{1}{2}} \left( \chi_{\uparrow \downarrow}(t) e^{\frac{i t}{\hbar}H_{AF}\left( \left< S_1^z-S_2^z \right>=1 \right)} \left| \uparrow \downarrow, n \right> \right. \nonumber \\
 && + \left. \chi_{\downarrow \uparrow }(t) e^{\frac{i t}{\hbar}H_{AF}\left( \left< S_1^z-S_2^z \right>=-1 \right)} \left| \downarrow \uparrow, n \right> \right).
\label{vec2}
\end{eqnarray}
In this equation $\left|n\right>$ represents the antiferromagnetic eigenstate of $H_{LM}$, and the time evolution is given for each component separately by the qubit mean field which corresponds to that specific component. This way the dynamics of the system do not get lost in the mean field approximation. With this expression for the time dependent eigenstates of the coupled system we can, as before, construct a density matrix and trace away all of the states of the antiferromagnet. The resulting reduced density matrix describes the decoherence of the qubit due to the coupling to the antiferromagnet, and the coherence time can be read off by looking at the off diagonal matrix element. We find that in the limit $\Delta \ll \gamma$ the qubit decoheres after a time $t_{coh}$ given by equation~(\ref{eq:t_coh}) --the same result as for the case where $\Delta =0$.

We thus conclude that decoherence occurs if a qubit is interacting with the orderparameter of a many-particle, symmetry broken manipulation device. The decoherence is caused by the energy shifts in the thin spectrum of the manipulation device, which are induced by the qubit. This thin spectrum is a generic feature that all quantum systems with a broken continuous symmetry share.

We thank Jan Zaanen and Mikhail Katsnelson for many fruitful suggestions and ongoing discussions.

\end{document}